\documentclass[twocolumn,showpacs,preprintnumbers,amsmath,amssymb]{revtex4}
\usepackage{graphicx}
\usepackage{dcolumn}
\usepackage{bm}
\usepackage{mathptmx}
\begin{document}
\title{Parametrization of the Driven Betatron Oscillation}
\author{R. Miyamoto}
\author{S. E. Kopp}
\affiliation{Department of Physics \\
             University of Texas at Austin \\
             Austin, Texas 78712 USA}
\author{A. Jansson}
\author{M. J. Syphers}
\affiliation{Fermi National Accelerator Laboratory \\
             Batavia, Illinois 60510 USA}
\date{\today}
\begin{abstract}
  An AC dipole is a magnet which produces a sinusoidally oscillating dipole field and excites
  coherent transverse beam motion in a synchrotron. By observing this coherent motion, the optical
  parameters can be directly measured at the beam position monitor locations. The driven
  oscillation induced by an AC dipole will generate a phase space ellipse which differs from that of
  the free oscillation. If not properly accounted for, this difference can lead to a
  misinterpretation of the actual optical parameters, for instance, of 6\% or more in the cases of
  the Tevatron, RHIC, or LHC. The effect of an AC dipole on the linear optics parameters is identical
  to that of a thin lens quadrupole. By introducing a new amplitude function to describe this new
  phase space ellipse, the motion produced by an AC dipole becomes easier to interpret. Beam
  position data taken under the influence of an AC dipole, with this new interpretation in mind, can
  lead to more precise measurements of the normal Courant-Snyder parameters. This new
  parameterization of the driven motion is presented and is used to interpret data taken in the FNAL
  Tevatron using an AC dipole. 
\end{abstract}
\pacs{41.85.-p, 29.27.-a}
\maketitle
\section{INTRODUCTION}
\begin{figure}
  \centering
  \includegraphics[width=55.0mm,clip]{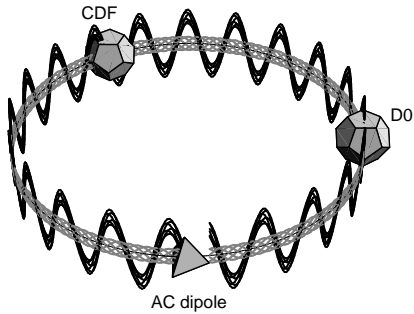}
  \caption{A diagram of the incoherent free oscillations (gray) and excited coherent oscillations
           (black) in the Tevatron. Since the oscillations of individual particles within the beam
           are incoherent, coherent oscillations must be excited to observe betatron motion and
           measure optical parameters. An AC dipole is a tool to excite sustained coherent
           transverse oscillations.}
  \label{fig:coherent.oscill.in.tev}
\end{figure}

  An AC dipole produces a sinusoidally oscillating dipole magnetic field and excites coherent
  transverse beam motion in a synchrotron for machine diagnosis (Fig
  \ref{fig:coherent.oscill.in.tev}). Unlike a conventional single turn kicker/pinger magnet, it
  drives the beam close to the betatron frequency typically for several thousands of revolutions. If
  the amplitude of its oscillating magnetic field is adiabatically ramped up and down, it can create
  large coherent oscillations without decoherence and emittance growth \cite{bai97}. This property
  makes it a useful diagnosis tool for a proton synchrotron, especially when it is used with an
  adequate beam position monitor (BPM) system. 

  AC dipoles have been used in the BNL RHIC \cite{bai01,bai02} and were also tested in the BNL AGS
  \cite{bai97} and CERN SPS \cite{berrig,schmidt}. Facilitated by its recently upgraded BPM system
  \cite{wolbers} (now with a resolution of 20 $\mu$m), a vertical AC dipole has been used in the
  FNAL Tevatron \cite{biw,pacopt,pacsys}. There is an ongoing project to develop AC dipoles
  for LHC as well.

  When the beam is driven by an AC dipole, the beam motion is governed by two driving terms and the
  influence of the lesser driving term makes the driven oscillation different from the free
  oscillation. Although this difference has typically been ignored in previous analyses
  \cite{bai02,peggs}, it could affect the interpretation of the linear optics more than 12\% in the
  Tevatron and 6\% in the RHIC and LHC.
  
  This paper proceeds as follows. Section II discusses the two driving terms produced by an AC
  dipole and presents a new formulation of the driven motion which is suited to treat the two
  driving terms at the same time. By introducing a new amplitude function $\beta_d$ for the driven
  motion, the difference between the free and driven oscillations becomes clear. Section III
  discusses the difference between the ordinary amplitude function for the free oscillation $\beta$
  and the newly defined amplitude function for the driven oscillation $\beta_d$. It is shown that
  the AC dipole has an analogy with a gradient error and, relative to $\beta$, $\beta_d$ behaves as
  if there is a gradient error. Section IV presents a few properties of the driven motion which were
  observed in the Tevatron. The new formulation matches the observed data well.
\section{A MODEL OF THE DRIVEN OSCILLATION}
\subsection{Two Driving Terms of an Oscillating Dipole Field}
\begin{figure}
  \centering
  \includegraphics[width=82.5mm,clip]{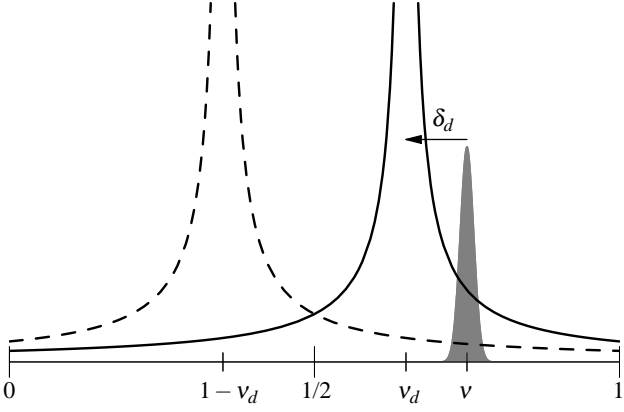}
  \put(-235,-8){0}
  \put(-122.5,-8){1/2}
  \put(-4.5,-8){1}
  \put(-161,-8){$1-\nu_d$}
  \put(-86,-8){$\nu_d$}
  \put(-63,-8){$\nu$}
  \put(-73,95){$\delta_d$}
  \caption{The amplitude of the driven motion versus the machine tune. A circulating beam is
           influenced by both (solid and dashed) of the resonant amplitudes. In typical operations
           of an AC dipole, $\delta_d$ is the order of 0.01 so that the primary driving tune $\nu_d$
           is outside of the tune spread (shaded area).}
  \label{fig:tune.spectrum}
\end{figure}
  The tune of an AC dipole $\nu_\text{acd}$ is defined as the ratio between the frequencies of the
  AC dipole $f_\text{acd}$ and the beam revolution $f_\text{rev}$: $\nu_\text{acd} = f_\text{acd}
  /f_\text{rev}$. In the following, for any tunes, only their fractional parts are considered. For
  instance, if $f_\text{acd}/f_\text{rev}$ is larger than one, $\nu_\text{acd}$ means the fractional
  part of $f_\text{acd}/f_\text{rev}$. Since the beam sees an AC dipole only once per revolution,
  the tune of an AC dipole $\nu_\text{acd}$ and $1-\nu_\text{acd}$ are equivalent (cf. Nyquist
  sampling theorem). Hence, under the influence of an oscillating dipole field a beam is driven by a
  pair of driving terms at $\nu_\text{acd}$ and $1-\nu_\text{acd}$. Obviously, the driving term
  closer to the machine tune $\nu$ ($0<\nu<1$) has bigger effects on a beam. In the following, the
  driving term closer to $\nu$ is called the primary and the other is called the secondary. A symbol
  $\nu_d$ is used for the primary driving tune:
\begin{align}
  \nu_d \equiv
  \begin{cases}
    \nu_\text{acd}   &\text{when} \quad |\nu_\text{acd}-\nu|     < |(1-\nu_\text{acd})-\nu| \\
    1-\nu_\text{acd} &\text{when} \quad |(1-\nu_\text{acd})-\nu| < |\nu_\text{acd}-\nu| ~.
  \end{cases}
\end{align}
  For example, the frequencies of the AC dipole and beam revolution in the Tevatron are
  $f_\text{acd} \simeq 20.5$ kHz and $f_\text{rev} \simeq 47.7$ kHz, respectively, and hence the
  tune of the AC dipole is $\nu_\text{acd} = 20.5/47.7 \simeq 0.43$. Since the machine tune of the
  Tevatron is $\nu \simeq 0.58$, $1-\nu_\text{acd} \simeq 0.57$ is the primary driving tune and
  $\nu_\text{acd} \simeq 0.43$ is secondary in this case (Table \ref{table:parameters}).

  The distance from the primary driving term to the machine tune $\delta_d \equiv \nu_d - \nu$ is an
  important parameter of the driven betatron oscillation. As seen later, the secondary driving term
  generates a difference between the free and driven oscillations and affects linear optics
  measurements. Ideally, if the beam is driven very close to the machine tune $\nu$ ($\delta_d
  \to 0$), the influence of the primary driving term becomes dominant and the secondary driving term
  can be ignored. In reality, however, the finite tune spread of the beam causes beam losses if
  $|\delta_d|$ is too small and there is always a lower limit for $|\delta_d|$
  (Fig \ref{fig:tune.spectrum}). AC dipoles are currently used in the Tevatron and RHIC and planned
  for the LHC. In these synchrotrons, the lower limit of $|\delta_d|$ is about 0.01 to prevent beam
  losses.

  When the amplitude of the field is constant, the position of the driven beam $x_d$ is given by
  \cite{peggs,tomas}
\begin{widetext}
\begin{align}
  x_d(nC+{\Delta}s)
  &\simeq \frac{\theta_\text{acd}\sqrt{\beta_\text{acd}}}{4\sin[\pi(\nu_\text{acd}-\nu)]}
          \sqrt{\beta({\Delta}s)}\cos[2\pi\nu_\text{acd}n+\psi({\Delta}s)
                                      +\pi(\nu_\text{acd}-\nu)+\chi_\text{acd}] \notag\\
  &\quad  +\frac{\theta_\text{acd}\sqrt{\beta_\text{acd}}}{4\sin[\pi((1-\nu_\text{acd})-\nu)]}
          \sqrt{\beta({\Delta}s)}\cos[2\pi(1-\nu_\text{acd})n+\psi({\Delta}s)
                                      +\pi((1-\nu_\text{acd})-\nu)-\chi_\text{acd}] ~,              \label{eq:xd0}
\end{align}
\end{widetext}
  where $C$ is the circumference of a ring, ${\Delta}s$ $(0\le{\Delta}s<C)$ is the longitudinal
  position measured from the location of the AC dipole, $\theta_\text{acd}$ is the maximum kick
  angle of the AC dipole, $\beta_\text{acd}$ is the amplitude function at the location of the AC
  dipole, $\psi$ is the phase advance of the free oscillation measured from the location of the AC
  dipole, and $\chi_\text{acd}$ is the initial phase of the AC dipole. The two terms in Eq
  \ref{eq:xd0} are completely symmetric and represent the influences of the two driving terms
  \footnote{Eq \ref{eq:xd0} is assuming the amplitude of the AC dipole field is adiabatically ramped
  up to a constant value. The exact expression of $x_d$ includes transient modes which are inversely
  proportional to the ramp up time and oscillate with the machine tune $\nu$. If the ramp up is slow
  enough, all of these modes are very small and decohere before the end of the ramp up. Hence these
  ignored modes do not affect the motion of the beam centroid but they may affect the beam size
  \cite{tomas}.}. To quantify the effect of the secondary driving term, it is useful to define a
  parameter which describes the ratio between the primary (larger) and secondary (smaller) modes in
  Eq \ref{eq:xd0}:
\begin{align}
  \lambda_d(\delta_d)
  \equiv \frac{\sin[\pi(\nu_d-\nu)]}{\sin[\pi((1-\nu_d)-\nu)]}
  =      \frac{\sin(\pi\delta_d)}{\sin(2\pi\nu+\pi\delta_d)} ~.                                     \label{eq:lambda}
\end{align}
  When $|\delta_d|=0.01$, $|\lambda_d| \simeq 0.06$ for the Tevatron with $\nu \simeq 0.58$ and
  $|\lambda_d| \simeq 0.03$ for the RHIC and LHC with $\nu \simeq 0.3$ and $0.7$
  (Table \ref{table:parameters}). This is the effect of the secondary driving term on the amplitude
  of the driven oscillation. When the machine tune is closer to the half-integer, the two driving
  terms are closer to each other and the influence of the secondary driving term gets larger. This
  is why $|\lambda_d|$ of the Tevatron is larger than that of the RHIC and LHC.
\begin{table}[h]
  \caption{Parameters related to the driven oscillation in the Tevatron, RHIC, and LHC when
           $|\delta_d|=0.01$. The secondary driving term affects the amplitude of the driven motion
           by $|\lambda_d|$ and, as seen later, produces an effect like $\beta$-beat with the 
           amplitude of $2|\lambda_d|$.}
  \begin{ruledtabular}
  \begin{tabular}{cccc}
    Parameter                       & Tevatron & RHIC  & LHC   \\
  \hline
    Machine Tune $\nu$              & .58      & .7    & .3    \\
    AC Dipole Tune $\nu_\text{acd}$ & .42      & .7    &       \\
    $|\lambda_d|$                   & 6\%      & 3\%   & 3\%   \\
    Amplitude of the $\beta$-beat   & 12-13\%  & 6-7\% & 6-7\% \\
  \end{tabular}
  \end{ruledtabular}
  \label{table:parameters}
\end{table}
\subsection{A New Parametrization of the Driven Betatron Oscillation}                               
  Eq \ref{eq:xd0} can be written in the following compact form which includes the influences of both
  driving terms:
\begin{align}
  x_d(s;\delta_d)
  = A_d(\delta_d)\sqrt{\beta_d(s;\delta_d)}\cos(\psi_d(s;\delta_d)\pm\chi_\text{acd}) ~.            \label{eq:xd}
\end{align}
  Here, $A_d$ is a quantity with dimensions of $(\text{length})^{1/2}$:
\begin{align}
  A_d(\delta_d) = \frac{\theta_\text{acd}}{4\sin(\pi\delta_d)}
                  \sqrt{(1-\lambda_d(\delta_d)^2)\beta_\text{acd}} ~,                               \label{eq:Ad}
\end{align}
  $\beta_d$ is a newly defined amplitude function of the driven oscillation which satisfies
\begin{align}
  \frac{\beta_d(s;\delta_d)}{\beta(s)}
  = \frac{1+\lambda_d(\delta_d)^2-2\lambda_d(\delta_d)\cos(2\psi(s)-2\pi\nu)}
         {1-\lambda_d(\delta_d)^2} ~,                                                               \label{eq:betad}
\end{align}
  $\psi_d$ is a newly defined phase advance of the driven oscillation measured from the location of
  the AC dipole:
\begin{align}
  \psi_d(s;\delta_d) = \int_0^{s} \!\! \frac{d\bar{s}}{\,\beta_d(\bar{s};\delta_d)} ~,
\end{align}
  and the sign in front of $\chi_\text{acd}$ is positive when $\nu_d = \nu_\text{acd}$ and negative
  when $\nu_d = 1-\nu_\text{acd}$. Hence, the driven oscillation can be parametrized in the same
  form as the free oscillation even when the influences of the both driving terms are included.
  Since $A_d$ is a constant of motion, the difference between the free and driven oscillations comes
  from the amplitude function $\beta_d$ and phase advance $\psi_d$. As discussed previously, in the
  limit, $\nu_d\to\nu$, the primary driving term becomes dominant and the secondary driving term can
  be ignored. In this limit $\lambda_d \to 0$ and $\beta_d$ and $\psi_d$ converge to $\beta$ and
  $\psi$.

  If the lesser mode in Eq \ref{eq:xd0} is ignored, the oscillation phase has an apparent jump of
  $2\pi\delta_d$ at the location of the AC dipole. However, if the influences of both driving terms
  are properly included as Eq \ref{eq:xd}, the phase advance is smooth at the location of the AC
  dipole. A relation between the phase advances of free and driven oscillations, $\psi$ and
  $\psi_d$, is given by 
\begin{align}
  \tan(\psi_d-\pi\nu_d)
  &= \frac{\,1+\lambda_d}{\,1-\lambda_d}\tan(\psi-\pi\nu) \notag\\
  &= \frac{\,\tan(\pi\nu_d)}{\tan(\pi\nu)}\tan(\psi-\pi\nu) ~.
\end{align}
  For the free oscillation, the phase advance in a single revolution is $\psi(s+C)-\psi(s)=2\pi\nu$
  (mod $2\pi$). In the equation above, $\psi_d = 2\pi\nu_d$ when $\psi = 2\pi\nu$. Hence, the phase
  advance in a single revolution is $2\pi\nu_d$ for the driven motion.
\section{DIFFERENCE BETWEEN THE AMPLITUDE FUNCTIONS $\beta$ and $\beta_d$}
  As seen in the previous section, the difference between the free and driven oscillations lies in
  the difference of their amplitude functions, $\beta$ and $\beta_d$. It is crucial to understand
  this difference between $\beta$ and $\beta_d$ in detail when an AC dipole is used to diagnose a
  synchrotron.

  In free betatron oscillations tune and amplitude function are coupled, and a change in tune
  involves a change in amplitude function and vice versa. This is true for the driven betatron
  oscillation, too. As seen in the previous section, for the driven oscillation, both the amplitude
  function and tune, $\beta_d$ and $\nu_d$, are different from those for the free oscillation,
  $\beta$ and $\nu$. As a matter of fact, the relation between these changes of the tune and
  amplitude function is the same as that for a gradient error. Hence, reviewing the effect of a
  gradient error is helpful to understand the driven oscillation.
\subsection{Review of a Gradient Error}
  If a synchrotron has a gradient error, its machine tune $\nu$ and amplitude function $\beta$
  change \cite{books}.
  Suppose a synchrotron has a gradient error with the strength $q_\text{err} = B'\ell/(B\rho)$ at
  the longitudinal position $s=0$. Then, the equation of motion is given by 
\begin{align}
  x''+k(s)x = -q_\text{err}\left[\sum_{n=-\infty}^\infty\!\!\delta(s-Cn)\right]x ~,                 \label{eq:eomq}
\end{align}
  where the prime denotes the derivative with the longitudinal coordinate $s$, $k$ is the spring
  constant, and $\delta$ is the Dirac's delta function.

  By comparing the single turn transfer matrices with and without the gradient error, the new
  tune $\nu_q$ and amplitude function $\beta_q$ satisfy the following two equations \cite{books}:
\begin{align}
  q_\text{err}
  &= 2\frac{\cos(2\pi\nu)-\cos(2\pi\nu_q)}{\beta_\text{err}\sin(2\pi\nu)}                           \label{eq:qq}\\
  \frac{\beta_q}{\beta}
  &= \frac{\sin(2\pi\nu)}{\sin(2\pi\nu_q)}
   - q_\text{err}\beta_\text{err}\frac{\sin\psi\sin(2\pi\nu-\psi)}{\sin(2\pi\nu_q)} ~,
\end{align}
  where $\beta_\text{err}$ is the amplitude function at the gradient error and $\psi$ is the phase
  advance measured from the gradient error. By substituting the first equation into the second, the
  ratio between the new and original amplitude functions $\beta_q/\beta$ is given by
\begin{align}
  \frac{\beta_q}{\beta}
  = \frac{1+\lambda_q^2-2\lambda_q\cos(2\psi-2\pi\nu)}{1-\lambda_q^2} ~.                            \label{eq:betaq}
\end{align}
  Here, $\lambda_q$ is a parameter similar to $\lambda_d$ in Eq \ref{eq:lambda}:
\begin{align}
 \lambda_q \equiv \frac{\sin(\pi\delta_q)}{\sin(2\pi\nu+\pi\delta_q)} ~,
\end{align}
  where $\delta_q$ is the tune shift by a gradient
  error $\delta_q \equiv \nu_q-\nu$. When the gradient error $q_\text{err}$ is small, the new and
  original amplitude functions satisfy
\begin{align}
  \frac{\beta_q-\beta}{\beta} \simeq -2\lambda_q\cos(2\psi-2\pi\nu) ~.                              \label{eq:beta.beat}
\end{align}
  This quantity behaves like a standing wave in a synchrotron and is called the $\beta$-beat (or
  sometimes $\beta$-wave). The amplitude of the $\beta$-beat is $2|\lambda_q|$.
\subsection{Analogy to a Gradient Error}
  As seen in Eqs \ref{eq:betad} and \ref{eq:betaq}, the relation between $\beta_d$ and $\delta_d$
  for an oscillating dipole field is the same as the relation between $\beta_q$ and $\delta_q$ for a
  gradient error. The following argument gives insight why an oscillating dipole field changes the
  observed phase space motion as like a gradient error.

  When the oscillation amplitude of the AC dipole field is constant, the Hill's equation of motion
  is given by
\begin{align}
  x''+k(s)x = -\sum_n\theta_\text{acd}\cos(2\pi{\nu_d}n\pm\chi_\text{acd})\delta(s-Cn) ~.           \label{eq:eomd0}
\end{align}
  The right-hand-side describes the kicks by the AC dipole located at $s=0$. The summation runs over
  the time period when the oscillation amplitude of the AC dipole field is constant and the sign in
  front of the initial phase $\chi_\text{acd}$ is the same convention as Eq \ref{eq:xd}. Eq
  \ref{eq:xd} is the particular solution of this inhomogeneous Hill's equation when the oscillation
  amplitude of the AC dipole field is adiabatically ramped to a constant amplitude. Since the phase
  of the driven oscillation $\psi_d$ increases by $2\pi\nu_d$ (mod $2\pi$) in one revolution, the
  position of the driven oscillation at the location of the AC dipole $s=Cn$ is given by
\begin{align}
  x_d(Cn;\delta_d) = A_d(\delta_d)\sqrt{\beta_d(0;\delta_d)}\cos(2\pi{\nu_d}n\pm\chi_\text{acd}) ~. \label{eq:xdacd}
\end{align}
  Notice the phases of the driven oscillation $x_d$ and the kicks by the AC dipole in Eq
  \ref{eq:eomd0} are both $2\pi{\nu_d}n\pm\chi_\text{acd}$ at the location of the AC dipole. Hence,
  when the beam passes the AC dipole, its magnetic field is proportional to the position of the
  driven oscillation $x_d$ like a quadrupole magnet. This is the physical reason why an oscillating
  dipole field changes the amplitude function like a gradient error. The phases of the driven
  oscillation and the AC dipole are synchronized like this only when the oscillation amplitude of
  the AC dipole field is constant after the adiabatic ramp up. Since $x_d$ is the solution of Eq
  \ref{eq:eomd0}, it formally satisfies the following equation
\begin{align}
  x''_d+k(s)x_d = -q_\text{acd}\left[\sum_n\delta(s-Cn)\right]x_d ~.                                \label{eq:eomd}
\end{align}
  Here, Eq \ref{eq:xdacd} is used to change the right-hand-side and $q_\text{acd}$ is a constant
  given by
\begin{align}
  q_\text{acd} = \frac{\theta_\text{acd}}{A_d\sqrt{\beta_d(0;\delta_d)}}
               = 2\frac{\cos(2\pi\nu)-\cos(2\pi\nu_d)}{\beta_\text{acd}\sin(2\pi\nu)} ~.            \label{eq:qd}
\end{align}
  Eq \ref{eq:eomd} is exactly the same as the Hill's equation with a gradient error, Eq
  \ref{eq:eomq}. By comparing Eqs \ref{eq:eomq}, \ref{eq:qq}, \ref{eq:qd}, and  \ref{eq:eomd}, it is
  trivial that the relation between $\beta_d$ and $\delta_d$ is the same as the relation between
  $\beta_q$ and $\delta_q$.
\subsection{Ring-wide Behavior of $\beta_d$}
  As discussed in the previous two sections, for the driven motion, the observed amplitude function
  $\beta_d$ differs from the actual $\beta$ as if there is a gradient error. Hence, $\beta_d$ is
  beating relative to $\beta$ and the beating amplitude is about $2|\lambda_d|$ from Eq
  \ref{eq:beta.beat}. Remember the effect of the secondary driving term on the beam motion is the
  order of $\lambda_d$. Since the amplitude function is proportional to the square of the position,
  its effect on the amplitude function is of the order of $2|\lambda_d|$. Since the minimum difference
  between the primary driving tune and machine tune $|\delta_d|$ is about 0.01 for the Tevatron,
  RHIC, and LHC, the beating amplitude of $\beta_d$ relative to $\beta$ is 12-13\% for the Tevatron
  and 6-7\% for the RHIC and LHC (Table \ref{table:parameters}).

  When turn-by-turn beam positions at all BPMs are given for the free oscillation, the relative
  $\beta$-function can be determined by simply comparing the square of the oscillation amplitude at
  each BPM. If the same analysis is applied to the turn-by-turn data of the driven oscillation, what
  is calculated is $\beta_d$ instead of $\beta$. If the difference between $\beta_d$ and $\beta$ is
  simply ignored and $\beta$ is determined in this way, the error may be as large as $2|\lambda_d|$.
  Furthermore, since the beating of $\beta_d$ cannot be distinguished from the real $\beta$-beat
  caused by gradient errors, the real $\beta$-beat cannot be measured in this way without depending
  on a machine model.

\begin{figure}
  \centering
  \includegraphics[width=85mm,clip]{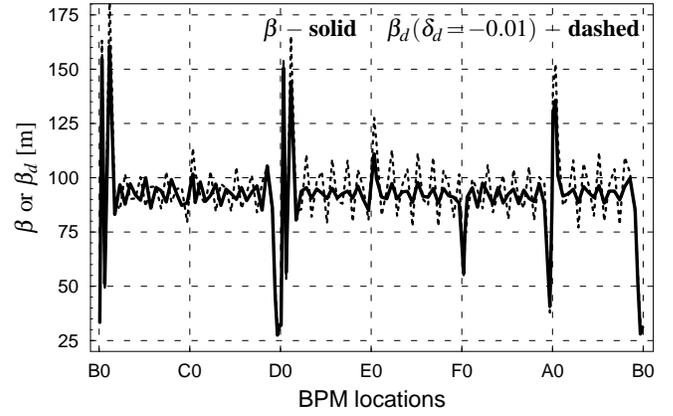}
  \put(-243,68){\rotatebox{90}{$\beta$ or $\beta_d$ [m]}}
  \put(-149,143){{\bf $\beta$ $-$ solid \quad $\beta_d(\delta_d\!=\!-0.01)$ $-$ dashed}}
  \caption{The amplitude functions of the free and driven oscillations, $\beta$ (solid) and
           $\beta_d$ when $\delta_d = -0.01$ (dashed). Both of them are calculated from
           turn-by-turn data of the driven oscillation. As expected, $\beta_d$ shows the 10-15\%
           beating relative to $\beta$. If the difference of $\beta$ and $\beta_d$ is simply
           ignored, the $\beta$ measurement has this much error and the real $\beta$-beat cannot
           be distinguished from the beating of $\beta_d$.}
  \label{fig:beta.vs.betad}
\end{figure}

  To calculate the true $\beta$-function from turn-by-turn data of the driven oscillation without
  depending on a machine model, multiple sets of data are necessary \cite{pacopt}. Fig
  \ref{fig:beta.vs.betad} shows amplitude functions of the free and driven oscillations, $\beta$ and
  $\beta_d(\delta_d = -0.01)$. They are both measured from data of the driven oscillation. Multiple
  data sets are used to calculate $\beta$ as described in \cite{pacopt} and $\beta_d$ is calculated
  by comparing the square of the amplitude at each BPM. As expected, $\beta_d$ is showing the beating
  of 10-15\% relative to $\beta$.
\subsection{Relation between $\beta_d$ and $\delta_d$}
\begin{figure*}
  \centering
  \includegraphics[width=178mm,clip]{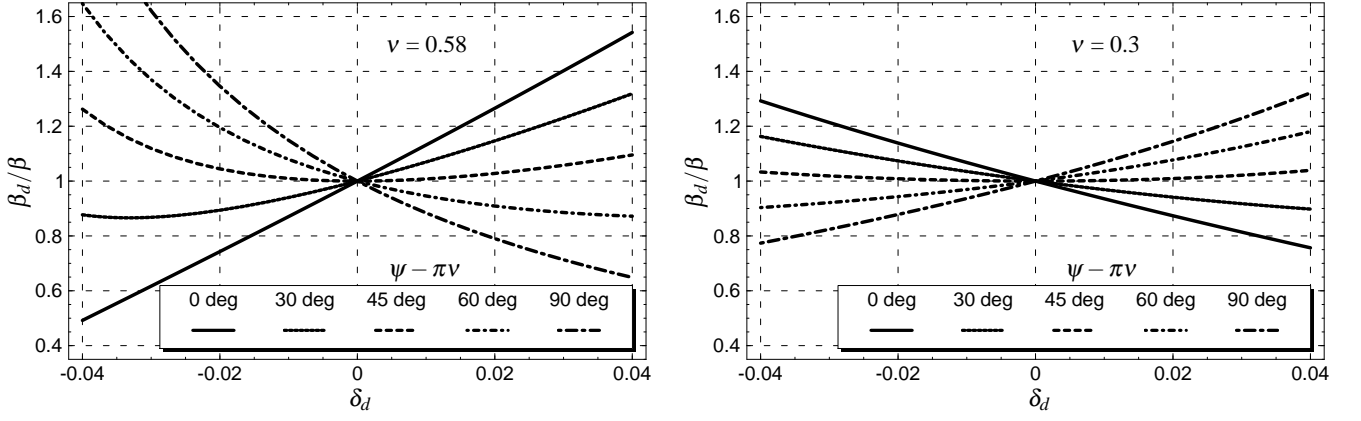}
  \put(-359,129){$\nu$ = 0.58}
  \put(-373,-5){$\delta_d$}
  \put(-502,70){\rotatebox{90}{$\beta_d/\beta$}}
  \put(-358,44){$\psi-\pi\nu$}
  \put(-100,129){$\nu$ = 0.3}
  \put(-116,-5){$\delta_d$}
  \put(-246,70){\rotatebox{90}{$\beta_d/\beta$}}
  \put(-102,44){$\psi-\pi\nu$}
  \caption{The relation between the amplitude functions of the free and driven betatron oscillations
           $\beta$ and $\beta_d$. The ratio $\beta_d/\beta$ is numerically calculated based on Eq
           \ref{eq:betad} by changing the difference between the primary driving tune and machine
           tune $\delta_d$ and the phase advance $\psi$. The left plot is when the machine tune 0.58
           like the Tevatron and the right is when 0.3 like the RHIC and LHC. Since the secondary
           driving term gets closer and $\lambda_d$ gets larger when the machine tune is closer to
           the half integer, $\beta_d/\beta$ is larger and the nonlinearity is stronger in the left
           plot. The nonlinearity gets larger when $\psi-\pi\nu$ gets closer to 45 deg and
           $\cos(2\psi-2\pi\nu)$ gets closer to zero.}
  \label{fig:betad.vs.delta_sim}
\end{figure*}
  The previous section discussed the global behavior of $\beta_d$ compared to $\beta$. This section
  considers how $\beta_d$ changes depending on $\delta_d$ at one location of a synchrotron. From Eq
  \ref{eq:betad}, the relation between $\beta_d$ and $\beta$ becomes nonlinear when $\lambda_d$ is
  large or the phase term $\cos(2\psi-2\pi\nu)$ is close to zero. Since the difference between
  $\beta$ and $\beta_d$ has a considerable impact on the linear optics measurement, it is important
  to understand the properties of Eq \ref{eq:betad} over wide ranges of parameters. Fig
  \ref{fig:betad.vs.delta_sim} shows the numerical calculations of $\beta_d/\beta$ based on Eq
  \ref{eq:betad}. The two plots are for two different machine tunes: $\nu = 0.58$ like the Tevatron
  and $\nu = 0.3$ like the RHIC and LHC. Since $\lambda_d$ is almost twice as large for the same
  $\delta_d$ when $\nu = 0.58$ compared to $\nu = 0.3$, the nonlinearity grows much faster with
  $\delta_d$ in the Tevatron. It is also seen in the left plot that the nonlinearity becomes larger
  when $|\cos(2\psi-2\pi\nu)|$ gets closer to zero. Such a nonlinear relation between $\beta_d$ and
  $\delta_d$ can be actually seen for the driven oscillation excited in the Tevatron. An example is
  shown in the next section.
\section{EVIDENCE OF THE SECONDARY DRIVING TERM}
\subsection{Rotation of the Phase Space Ellipse}
\begin{figure}[b]
  \centering
  \includegraphics[width=85.0mm,clip]{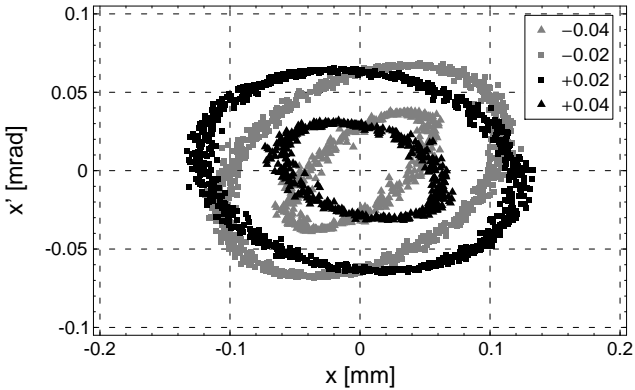}
  \caption{Measured phase space ellipses of the driven oscillations when $\delta_d =$ $\pm0.02$ and
           $\pm0.04$. Here, $\delta_d$ is the difference between the primary driving tune and the
           machine tune. The figure shows the phase space at one of the low-$\beta$ locations (B0)
           in the Tevatron where the derivative of the amplitude function $\alpha$ is zero by
           design. Since the Courant-Snyder-like parameters of the driven oscillation $\beta_d$,
           $\alpha_d$, and $\gamma_d$ depend on $\delta_d$, not only the area but also the shape of
           the ellipse changes with $\delta_d$.}
  \label{fig:phase.space}
\end{figure}
  The previous section discussed the amplitude function of the driven motion $\beta_d$. Parameters
  corresponding to the other Courant-Snyder parameters $\alpha$ and $\gamma$ can be also defined as
  for the free oscillation:
\begin{align}
  \alpha_d(s;\delta_d) &\equiv -\frac{1}{2}\frac{d\beta_d(s;\delta_d)}{ds} \\
  \gamma_d(s;\delta_d) &\equiv \frac{1+\alpha_d(s;\delta_d)^2}{\beta_d(s;\delta_d)} ~.
\end{align}
  The explicit forms of these parameters are given by
\begin{align}
  \alpha_d
  &= \frac{1+\lambda_d^2-2\lambda_d\cos(2\psi-2\pi\nu)}{1-\lambda_d^2}\,\alpha \notag\\                                                                           
  &  \hspace{95pt} -\frac{2\lambda_d\sin(2\psi-2\pi\nu)}{1-\lambda_d^2}
\end{align}
  and
\begin{align}
  \gamma_d
  = \frac{1+\lambda_d^2+2\lambda_d\cos(2\psi-2\pi\nu+2\arctan\alpha)}{1-\lambda_d^2}\,\gamma ~.
\end{align}
  When $\beta_d$, $\alpha_d$, $\gamma_d$, and $A_d$ are defined this way, they satisfy the
  Courant-Snyder invariance:
\begin{align}
  A_d^2 = \gamma_dx_d^2+2\alpha_dx_dx'_d+\beta_dx_d^{\prime\,2} ~.                                  \label{eq:ellipse}
\end{align}
  Hence, the turn-by-turn position and angle of the driven oscillation also form an ellipse on the
  phase space, like the free oscillation. Since not only $A_d$ but also the Courant-Snyder-like
  parameters $\beta_d$, $\alpha_d$, and $\gamma_d$ depend on the difference between the primary
  driving tune and the machine tune $\delta_d$, both the area and shape of the phase space ellipse
  changes with $\delta_d$ for the driven oscillation. In two collision straight sections of the
  Tevatron, B0 and D0, there are pairs of BPMs with no magnetic element in-between. The beam travels
  along straight lines between these pairs and, hence, both position and angle can be directly
  measured at these locations. Fig \ref{fig:phase.space} shows the measured phase ellipses of the
  driven oscillations by using a pair of such BPMs. The frequency of the AC dipole was changed to
  adjust $\delta_d$ to $\pm0.04$ and $\pm0.02$, while the kick angle of the AC dipole
  $\theta_\text{acd}$ was kept the same. As expected, the shape of the phase space ellipse changes
  with $\delta_d$. Since $\delta_d$ dependence of $\beta_d$, $\alpha_d$, and $\gamma_d$ comes from
  the secondary driving term, the rotation of the phase space ellipse is its qualitative evidence. 

\begin{figure}
  \centering
  \includegraphics[width=85mm,clip]{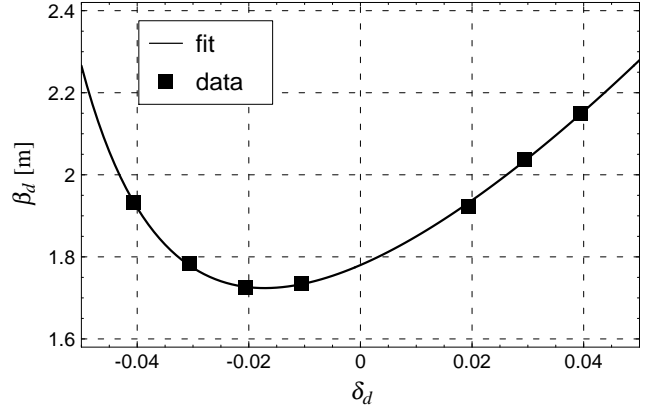}
  \put(-112,-10){$\delta_d$}
  \put(-240,63){\rotatebox{90}{$\beta_d$ [m]}}
  \caption{The relation between the amplitude function of the driven oscillation $\beta_d$ and the
           difference between the primary driving tune and the machine tune $\delta_d$. The location
           is the same low-$\beta$ point of the Tevatron (B0) in Fig \ref{fig:phase.space}. The
           amplitude function $\beta_d$ at each data point is determined from the shape of an
           ellipse in Fig \ref{fig:phase.space}. The curve is the fit of Eq \ref{eq:betad} to the
           data points. Despite the strong nonlinearity, Eq \ref{eq:betad} is fitting well. In the
           figure, $\beta_d(\delta_d=0)$ corresponds to the true value of $\beta$ at the location.}
  \label{fig:betad.vs.delta}
\end{figure}

  By fitting Eq (\ref{eq:ellipse}) to an ellipse in Fig \ref{fig:phase.space}, its area
  ${\pi}A_d^2$ and the parameters $\beta_d$, $\alpha_d$, and $\gamma_d$ can be determined. Fig
  \ref{fig:betad.vs.delta} shows $\beta_d$ determined from the fits to ellipses in Fig
  \ref{fig:phase.space} (and three more). The curve in the figure is the fit of Eq \ref{eq:betad} to
  the data with parameters $\beta$ and $\psi$. The model of Eq \ref{eq:betad} is fitting well to the
  data even though the nonlinearity is strong in the relation between $\beta_d$ and $\delta_d$ at
  the location. The $\beta$-function at the location can be calculated as one of the fit parameters.
  In the figure it is the value of $\beta_d$ when $\delta_d=0$.
\subsection{Asymmetric Amplitude Response}
\begin{figure}
  \centering
  \includegraphics[width=85.0mm,clip]{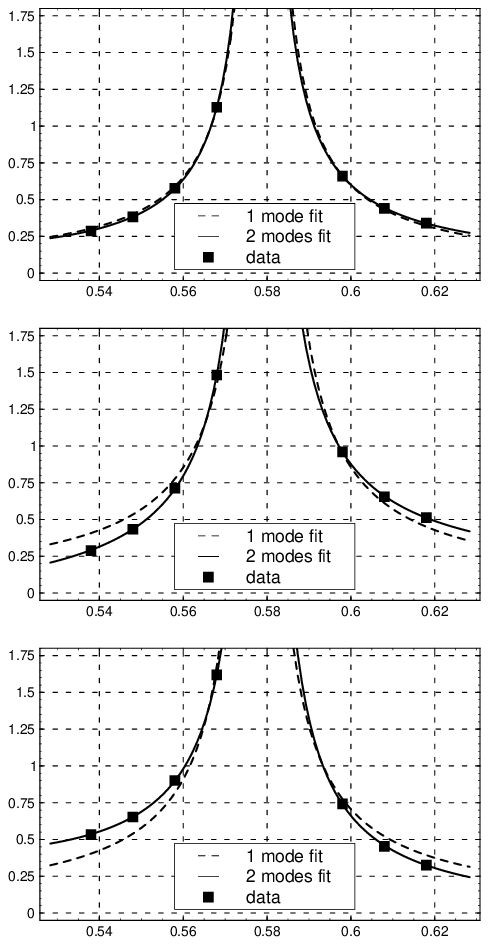}
  \put(-147,298){primary driving tune $\nu_d$}
  \put(-240,349){\rotatebox{90}{amplitude [mm]}}
  \put(-75,432){$\cos(2\psi-2\pi\nu)$}
  \put(-52,422){$\simeq 0.30$}
  \put(-147,147){primary driving tune $\nu_d$}
  \put(-240,199){\rotatebox{90}{amplitude [mm]}}
  \put(-75,281.5){$\cos(2\psi-2\pi\nu)$}
  \put(-52,271.5){$\simeq 0.88$}
  \put(-147,-3){primary driving tune $\nu_d$}
  \put(-240,49){\rotatebox{90}{amplitude [mm]}}
  \put(-75,131){$\cos(2\psi-2\pi\nu)$}
  \put(-59,121){$\simeq -0.97$}
  \caption{Measured relation between the amplitude of the driven oscillation and the primary driving
           tune $\nu_d$ at three BPM locations in the Tevatron. In the measurements, only $\nu_d$
           was changed. The solid and dashed lines are fits with and without the effect of the
           secondary driving term. The asymmetry around the tune $\nu \simeq 0.5785$ increases with
           $|\cos(2\psi-2\pi\nu)|$. In the second plot where $\cos(2\psi-2\pi\nu) > 0$, the
           amplitude is larger in the region $\nu_d > \nu$. As expected from Eq \ref{eq:amp2b}, the
           relation flips in the third plot where $\cos(2\psi-2\pi\nu) < 0$.}
  \label{fig:amp.vs.nud}
\end{figure}
  When the secondary driving term is negligible, by ignoring the smaller term of Eq \ref{eq:xd0} or
  taking the limit of $\lambda_d \to 0$ in Eqs \ref{eq:xd}, \ref{eq:Ad}, and \ref{eq:betad}, the
  amplitude of the driven oscillation can be approximated by
\begin{align}
  a_d^{(0)}(s;\delta_d)
  \equiv \frac{\theta_\text{acd}\sqrt{\beta_\text{acd}\beta}}{4|\sin(\pi\delta_d)|} ~.              \label{eq:amp1}
\end{align}
  In this case, the amplitude of the driven oscillation depends on the primary driving tune $\nu_d$
  only through $|\sin(\pi\delta_d)|$ (remember $\delta_d=\nu_d-\nu$) and is symmetric around the
  machine tune $\nu$. From Eqs \ref{eq:xd}, \ref{eq:Ad}, and \ref{eq:betad}, the amplitude
  including the effect of the secondary driving term $a_d(s;\delta_d)$ is given by
\begin{align}
  a_d(s;\delta_d) = a_d^{(0)}\sqrt{1+\lambda_d^2-2\lambda_d\cos(2\psi-2\pi\nu)} ~.                  \label{eq:amp2}
\end{align}
  Now, the amplitude depends on $\nu_d$ through the factor $[1+\lambda_d^2-2\lambda_d\cos(2\psi(s)
  -2\pi\nu)]^{1/2}$ as well. To the first order in $\delta_d$, the amplitude is approximated by
\begin{align}
  a_d \simeq a_d^{(0)}\left[1-\frac{\pi\cos(2\psi(s)-2\pi\nu)}{\sin(2\pi\nu)}\,\delta_d\right] ~.   \label{eq:amp2b}
\end{align}
  Hence, the secondary driving term makes the $\nu_d$ dependence of the amplitude asymmetric around
  the machine tune $\nu$. The magnitude of this asymmetry at each location is determined by the
  factor $\cos(2\psi-2\pi\nu)$.

  Fig \ref{fig:amp.vs.nud} shows the relation between the amplitude of the driven oscillation and
  $\nu_d$ at three BPM locations in the Tevatron. In the measurements, only the frequency of the AC
  dipole was changed while its kick angle $\theta_\text{acd}$ was kept the same. The dashed and
  solid lines represent the fits of Eq \ref{eq:amp1} and Eq \ref{eq:amp2} to the data. The fit
  parameters are $\theta_\text{acd}[\beta_\text{acd}\beta]^{1/2}$ and $\nu$ for Eq \ref{eq:amp1} and
  $\theta_\text{acd}[\beta_\text{acd}\beta]^{1/2}$, $\nu$, and $\psi$ for Eq \ref{eq:amp2}
  \footnote{The ring-wide $\beta$-function can be determined from the fit up to a constant
  $\theta_\text{acd}\sqrt{\beta_\text{acd}}$. The constant is determined from the analysis using a
  pair of BPMs in the collision straight sections. See \cite{pacopt} for details. The phase advance
  $\psi$ can be determined for the fit, too.}. At two locations where $|\cos(2\psi-2\pi\nu)|$ is
  close to one, the asymmetry around the machine tune ($\nu \simeq 0.5785$) is large and the result
  of the fits without the secondary driving term (Eq \ref{eq:amp1}) is not well matched.

\begin{figure}[h]
  \centering
  \includegraphics[width=85mm,clip]{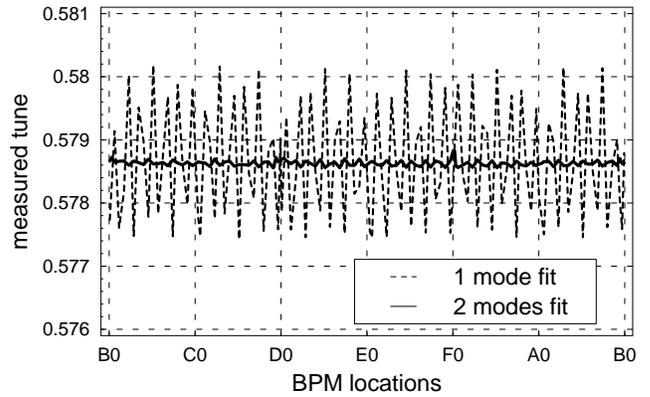}
  \caption{Measured machine tune at all BPM locations from the fits of the amplitude versus the
           primary driving tune in Fig \ref{fig:amp.vs.nud}. The solid line includes the influence
           of the secondary driving term and the dashed line does not. Since the tune is a global
           parameter of a synchrotron, the equation with the secondary driving term is the better
           model of the driven oscillation. The beating of the dashed line is caused by ignoring the
           beating of $\beta_d$ relative to $\beta$.}
  \label{fig:tune}
\end{figure}

  Although the existence of a secondary driven term effect is clear in Fig \ref{fig:amp.vs.nud},
  there is better evidence that Eq \ref{eq:amp2} fits the data better than \ref{eq:amp1}. From the
  fits in Fig \ref{fig:amp.vs.nud}, the machine tune $\nu$ can be determined at each BPM location.
  Fig \ref{fig:tune} shows machine tunes determined at all BPM locations from the fits of the
  amplitude versus $\nu_d$. The dashed and solid lines represent machine tunes from the fits of Eqs
  \ref{eq:amp1} and \ref{eq:amp2}. Since the machine tune $\nu$ is a global parameter of a
  synchrotron, the variation of the measured machine tune over BPMs shows the inaccuracy of the
  measurement. From the figure, it is clear the model including the secondary driving fits to the
  data much better. This also shows the importance of the secondary driving term in the driven
  oscillation.
\section{CONCLUSION}
  Under the influence of a sinusoidally oscillating magnetic field of an AC dipole, the beam is
  driven by two driving terms. As a result the phase space trajectory of the driven motion is
  different from that of the free betatron motion. If the difference is simply ignored,
  interpretation of the linear optics based on the data of the driven motion can have 12-13\% error
  for the Tevatron and 6-7\% error for the RHIC and LHC. This difference on the phase space is
  identical to the influence of a gradient error at the same location as the AC dipole. Hence, as a
  gradient error changes the amplitude function around the ring, the expression of the driven motion
  can be simplified by introducing a new amplitude function for the driven motion.

  This paper presented a few examples of the difference between the free and driven motions as
  observed in the Tevatron. It also showed that the new parametrization of the driven motion
  clarifies the data interpretation and multiple data sets are necessary to better resolve the true
  optical parameters of the free oscillation.
  
  With this knowledge, very precise and direct measurements of the amplitude function in a hadron
  synchrotron can be obtained quickly without degradation of the beam quality, using a small number
  of data sets obtained at different frequencies of the AC dipole. This technique will be
  especially useful in the LHC, for example, to adjust the beam envelope at critical locations such
  as at beam collimation devices.


\newpage
\begin{thebibliography}{99}
  \bibitem{bai97}  M. Bai {\it et al}., Phys. Rev. E {\bf 56}, p. 6002 (1997).
  \bibitem{bai01}  M. Bai {\it et al}., in {\it Proceedings of the 19th Particle Accelerator
                   Conference, Chicago, Illinois, 2001} (IEEE, Piscataway, NJ, 2001), p. 3606.
  \bibitem{bai02}  M. Bai {\it et al}., in {\it Proceedings of the 8th European Particle Accelerator
                   Conference, Paris, France, 2002} (EPS-IGA and CERN, Geneva, 2002), p. 1115.
  \bibitem{berrig} O. Berrig {\it et al}., in {\it Proceedings of the 5th European Workshop on Beam
                   Diagnostics and Instrumentation for Particle Accelerators, Grenoble, France,
                   2001} (ESRF, Grenoble, 2001), p. 82.
  \bibitem{schmidt}F. Schmidt {\it et al}., CERN Report No. AB-Note-2003-031 MD, 2003.
  \bibitem{wolbers}S. Wolbers {\it et al}., in {\it Proceedings of the 21st Particle Accelerator
                   Conference, Knoxville, Tennessee, 2005} (IEEE, Piscataway, NJ, 2005), p. 410.
  \bibitem{biw}    R. Miyamoto {\it et al}., in {\it Proceedings of the 12th Beam Instrumentation
                   Workshop, Batavia, Illinois, 2006} (AIP, Melville, New York, 2006), p. 402.
  \bibitem{pacopt} R. Miyamoto {\it et al}., in {\it Proceedings of the 22nd Particle Accelerator
                   Conference, Albuquerque, New Mexico, 2007} (IEEE, Piscataway, NJ, 2007), p. 3465.
  \bibitem{pacsys} R. Miyamoto {\it et al}., in {\it Proceedings of the 22nd Particle Accelerator
                   Conference, Albuquerque, New Mexico, 2007} (IEEE, Piscataway, NJ, 2007), p. 3868.
  \bibitem{peggs}  S. Peggs, in {\it Proceedings of the 18th Particle Accelerator Conference, New
                   York, New York, 1999} (IEEE, Piscataway, NJ, 1999), p. 1572.
  \bibitem{tomas}  R. Tomas, Phys. Rev. ST Accel. Beams {\bf 8}, 024401 (2005).
  \bibitem{books}  See standard textbooks of accelerator physics such as D. A. Edwards and M. J.
                   Syphers, {\it An Introduction to the Physics of High Energy Accelerators} (John
                   Wiley \& Sons, Inc., New York, 1993) or S. Y. Lee, {\it Accelerator Physics}
                   ({\it Second Edition}) (World Scientific, Singapore, 2004).
  \bibitem{syphers}M. J. Syphers and R. Miyamoto, in {\it Proceedings of the 22nd Particle
                   Accelerator Conference, Albuquerque, New Mexico, 2007} (IEEE, Piscataway, NJ,
                   2007), p. 3495.
\end{thebibliography}
\end{document}